\documentclass[twocolumn,aps,prb,showpacs]{revtex4}
\usepackage{amssymb}
\usepackage{graphicx}

\begin{document}

\title{Ostwald ripening of faceted two-dimensional islands}
\author{V. M. Kaganer}
\author{W. Braun}
\affiliation{Paul--Drude--Institut f\"{u}r Festk\"{o}rperelektronik,
Hausvogteiplatz 5--7, 10117 Berlin, Germany}
\author{K. K. Sabelfeld}
\affiliation{Weierstra\ss --Institut f\"ur Angewandte Analysis und
             Stochastic, Mohrenstr.\ 39, 10117 Berlin, Germany}
\affiliation{Institute of Computational Mathematics and Mathematical
             Geophysics, Russian Academy of Sciences,\\
             Lavrentiev Prosp.\ 6, 630090 Novosibirsk, Russia}
\begin{abstract}
We study Ostwald ripening of two-dimensional adatom and advacancy
islands on a crystal surface by means of kinetic Monte Carlo
simulations. At large bond energies the islands are square-shaped,
which qualitatively changes the coarsening kinetics. The
Gibbs--Thomson chemical potential is violated: the coarsening
proceeds through a sequence of `magic' sizes corresponding to square
or rectangular islands. The coarsening becomes attachment-limited,
but Wagner's asymptotic law is reached after a very long transient
time. The unusual coarsening kinetics obtained in Monte Carlo
simulations are well described by the Becker--D\"oring equations of
nucleation kinetics. These equations can be applied to a wide range
of coarsening problems.
\end{abstract}

\date{\today}



\maketitle

\section{Introduction}

Domains of a guest phase inside a matrix tend to coarsen, thus
reducing their specific interface energy. The prominent mechanism of
coarsening was proposed by Ostwald\cite{ostwald00} more than hundred
years ago: larger domains grow at the expense of smaller ones by
exchanging atoms. The net atom flux is directed to larger domains
since they possess smaller interface energy per atom. The seminal
theory of Ostwald ripening was proposed by Lifshitz and
Slyozov\cite{lifshitzslezov61} and by Wagner.\cite{wagner61} They
showed that, at late times, the system is characterized by a single
characteristic scale, namely, the average domain size $R(t)$. The
time evolution of the system consists in changing the scale: the
domain distribution, shape of the diffraction peaks, etc.\ remain
unchanged when scaled by $R(t)$. The average domain size follows, in
turn, universal laws, $R(t)\propto t^{1/3}$ if the atom diffusion is
the rate limiting process\cite {lifshitzslezov61} and $R(t)\propto
t^{1/2}$ if the attachment-detachment at the domain interface is the
limiting one.\cite{wagner61}

The kinetic scaling is essentially based on the Gibbs--Thomson
formula $\mu =\gamma /R$ for the excess chemical potential of a gas
that is in equilibrium at the curved surface of a liquid droplet
(the constant $\gamma $ is proportional to the surface tension). The
aim of the present work is to study the Ostwald ripening kinetics at
low temperatures (or large bond energies) when the crystalline
droplets are faceted. The energy of a small crystalline droplet is
minimum at `magic' sizes when all facets are completed. The
coarsening proceeds as a sequence of jumps from one magic size to
the next. We perform kinetic Monte Carlo simulations of Ostwald
ripening kinetics for faceted two-dimensional\ (2D) islands and find
very long transient behavior of the system, so that the universal
asymptotic laws are still not reached. We develop a mean-field
theory for Ostwald ripening, based on the
Becker--D\"{o}ring\cite{BeckerDoering35} equations. We show that
these equations, being the basic equations of nucleation
theory,\cite {frenkel46,LewisAnderson78} can be used to describe the
coarsening kinetics in the whole size range, starting from monomers
up to the long-time asymptotics that are not available in Monte
Carlo simulations. Both the Lifshitz--Slyozov--Wagner regime and the
coarsening through a sequence of magic sizes are well described.
This approach requires only the knowledge of the droplet energy
dependence on the number of atoms in the droplet and can be applied
to a wide range of coarsening problems in other systems as well.

Two-dimensional (2D) islands on a crystal surface are a practically
important physical system that reveals different coarsening
mechanisms and allows detailed theoretical and experimental studies
of the coarsening kinetics. From the experimental studies, we
mention the ones that report time exponents $n$ in the coarsening
law $R(t)\propto t^{n}$. These include low-energy electron
diffraction from a chemisorbed monolayer of oxygen on
W(110),\cite{tringides87,wu89} helium atom beam diffraction from 0.5
monolayer (ML) of Cu on Cu(100),\cite{ernst92} optical microscopy of
a thin layer of succinonitrile within the liquid-solid coexistence
region\cite {krichevsky93,krichevsky95} and a binary mixture of
amphiphilic molecules,\cite{seul94} and low-energy electron
microscopy of Si on Si(001).\cite {theis95,bartelt96} In these
works,\cite{tringides87,wu89,ernst92,krichevsky93,krichevsky95,seul94}
the time exponents somewhat smaller than $1/3$ were found and
explained by the Lifshitz--Slyozov law with finite-size corrections.
The time exponent $1/2$ obtained for Si on
Si(001)\cite{theis95,bartelt96} was treated as the case of kinetics
limited by the attachment and detachment of adatoms to
steps.\cite{wagner61} Our recent x-ray diffraction study of
coarsening of 2D GaAs islands on GaAs(001),\cite{braun04prb} which
showed an apparent time exponent close to 1, was the experimental
inspiration for the present work.

Two-dimensional islands of `magic' sizes were observed on several
surfaces, such as Pt(111)\cite{rosenfeld92},
Si(111),\cite{voigtlander98} and Ag(111)\cite{morgenstern05} (see
also a review\cite{wang01}). It was shown theoretically that the
presence of magic island sizes disrupts the scaling law of
submonolayer molecular beam epitaxy growth.\cite {schroeder95} Magic
sizes of three-dimensional Pb nanocrystals on Si(111) lead to a
breakdown of the classical Ostwald ripening laws.\cite{jeffrey06}

Monte Carlo simulations of Ostwald ripening were performed using the
2D Ising model.\cite{huse86,amar88,roland89} They were limited to
rather small values of the coupling constant, so that the domains
are rounded and faceting is absent. The time exponents were found to
be smaller than $1/3$, which was explained by finite-size
corrections to the Lifshitz--Slyozov law. Further discussion of
theoretical and simulation studies can be found in several
reviews.\cite {voorhees85,mouritsen90review,bray94} Despite kinetic
Monte Carlo simulations are routinely used to model epitaxial
growth,\cite{clarke87,clarke88,clarke89,clarke91,shitara92a,shitara92b}
we are aware of only one such study of coarsening of 2D islands on a
crystal surface.\cite{lam99} This latter simulation was limited to
small bond energies and rounded islands, similar to the simulations
of the Ising model.

A physical difference between the coarsening kinetics of 2D
epitaxial islands and that of Ising spins becomes evident when we
compare adatoms and advacancies on one side with up and down spins
on the other side. The first two objects possess qualitatively
different kinetics (motion of an advacancy is a result of the
collective motion of atoms), while up and down spins are equivalent.
This distinction manifests itself in the transition probabilities,
as discussed below. The fundamental laws of Ostwald ripening are
expected to be independent of the transition probability
distribution, so that a kinetic Monte Carlo simulation of the
coarsening of epitaxial islands allows one to check this conclusion.
Here, we perform kinetic Monte Carlo simulations of Ostwald ripening
of 2D adatom islands (surface coverage 0.1 ML)\ and 2D advacancy
islands (surface coverage 0.9 ML) in a wide range of bond energies
(or temperatures). Our particular aim is to perform simulations in
the case of large bond energies (low temperatures)\ when the islands
are faceted, which was not studied previously.

\section{Monte Carlo simulations}

\subsection{Simulation method}

We employ the well-established generic model developed for kinetic
Monte Carlo simulations of molecular beam epitaxy.\cite
{clarke87,clarke88,clarke89,clarke91,shitara92a,shitara92b,lam99}
Atoms occupy a simple cubic lattice and interact with a pair energy
that depends only on the number of bonds. An alternative approach to
simulate surface kinetics is a detailed Monte Carlo simulation of a
particular surface with energetic parameters taken from ab initio
calculations, as it was done for GaAs(001) or InAs(001).\cite
{kratzer99,kratzer02,grosse02prl,grosse02prb1,grosse02prb2} Such
simulations are very time-consuming and hence are limited to small
time and spatial scales. They can hardly be applied to study the
coarsening process. Some characteristic features of compound
semiconductors can, however, be included in the generic model as a
compromise.\cite{heyn97a,heyn97b,zhang03}

We use an algorithm\cite{bortz75} that advances simulated time
depending on the probability of the chosen event. This algorithm is
commonly used in the epitaxial growth simulations. We note that the
Ostwald ripening simulations of the 2D Ising
model\cite{huse86,amar88,roland89} have employed the Metropolis
accept--reject algorithm. This algorithm becomes ineffective at low
temperatures, since most of the attempts are rejected and computer
time is wasted. That is why previous
simulations\cite{huse86,amar88,roland89} were restricted to
relatively high temperatures $T>0.5T_{c}$, where $T_{c}$ is the
Ising phase transition temperature. Of course, both algorithms give
the same results and differ only in the computation time.

The choice of the probability $w(\mathbf{x}\rightarrow \mathbf{y})$
for the transition from the state $\mathbf{x}$ to the state
$\mathbf{y}$ incorporates the physics of the system into the
simulations. The choice is made differently for the epitaxial growth
and the Ising model simulations. It is worthwhile to compare these
probabilites briefly. A sufficient condition that the system evolves
to thermodynamic equilibrium is the detailed balance condition,
$w(\mathbf{x} \rightarrow \mathbf{y})/w(\mathbf{y}\rightarrow
\mathbf{x})=\exp (-\Delta E/k_{\mathrm{B}}T)$. Here $\Delta
E=E(\mathbf{y})-E(\mathbf{x})$ is the energy difference between the
states $\mathbf{x}$ and $\mathbf{y}$, $k_{\mathrm{B}}$ is the
Boltzmann constant and $T$ is the temperature. The simulations of
the Ising model use a probability that depends on $\Delta E$ (either
the Metropolis or the Glauber probability). These probabilities
favor transitions which reduce the energy of the system, $\Delta
E<0$. On the other hand, for an atom jump on the crystal surface,
the transition probability does not depend on the final state
$\mathbf{y}$ but only on the height of the energy barrier that needs
to be overcome.\cite{kang89} The probability is
$w(\mathbf{x}\rightarrow \mathbf{y})\propto \exp [E(\mathbf{x})
/k_{\mathrm{B}}T]$, where $E(\mathbf{x})<0$ is the energy of the
initial state with respect to the barrier. Such a probability
obviously satisfies the detailed balance condition. The system
evolves into a lower-energy state since it escapes higher-energy
initial states with larger probabilities.

In the present study, no step edge barrier is imposed. An atom
detaching from a step edge can go to the lower or the upper terrace
with equal probabilities. In particular, atom exchange between
advacancy islands is achieved predominantly by adatoms diffusing on
the top level rather than by the diffusion of vacancies, despite
that the latter process is not forbidden. Similar simulations, but
with an infinite step edge barrier, were performed in our preceding
work.\cite{kaganer06coalescence} In that study, the restriction for
atoms to escape a vacancy island to the higher level resulted in
another coarsening mechanism, diffusion and coalescence of whole
islands due to atom detachment and reattachment within an island.
The coarsening by dynamic coalescence is much less effective than
Ostwald ripening and becomes essential when the detachment of atoms
form islands is prohibited.

An atom that has $n$ neighbors in the initial state with equal bond
energies $E_{b}$ to these neighbors, possesses an energy
$E(\mathbf{x}) = -(nE_{b}+E_{D})$, where the activation energy of
surface diffusion $E_{D}$ is the barrier height. It determines the
time scale $\tau $ of the problem, $\tau ^{-1} = \nu \exp (-E_{D}
/k_{\mathrm{B}}T)$, where $\nu \approx 10^{13}$ s$^{-1}$ is the
vibrational frequency of atoms in a crystal. In the epitaxial growth
simulations, the time scale $\tau $ is to be compared with the
deposition flux, which determines an appropriate choice of $E_{D}$.
We do not consider deposition, and the choice of $E_{D}$ is
arbitrary. Note that the works on the Ising model kinetics measure
time simply in the flip attempts (sweeps) per lattice site. We take
the same values of $E_{D}$ as in the preceding
work,\cite{kaganer06coalescence} with the aim to compare time scales
of Ostwald ripening (in absence of the step edge barrier)\ with that
of dynamic coalescence (infinite step edge barrier). Namely, we take
$E_{D}=0.2$; $0.1$; $0$~eV for $E_{b}=0.2$; $0.3$; $0.4$~eV,
respectively.

The ratio of the interaction energy between neighboring atoms to the
temperature $E_{b}/k_{\mathrm{B}}T$ is the only essential parameter
for the coarsening problem. We fix the temperature at 400 K\ and
vary the bond energy $E_{b}$ from 0.2 eV to 0.4 eV. In terms of our
model, the Ising phase transition takes place at
$E_{b}/k_{\mathrm{B}}T=2\ln (1+\sqrt{2})$. Our choice of bond
energies corresponds to $T/T_{c}$ varying from 0.15 to 0.3,
temperatures much lower than the ones used in previous kinetic Monte
Carlo studies of Ostwald ripening.\cite
{huse86,amar88,roland89,lam99} Here $T_{c}$ is the Ising phase
transition temperature.

We perform kinetic Monte Carlo simulations on a 1000$\times $1000
square grid with periodic boundary conditions. Each simulation is
repeated 25 times, to obtain sufficient statistics for the island
size distribution. In the initial state, either 0.1 ML\ or 0.9 ML
are randomly deposited. Adatom islands form in the first case and
advacancy islands in the second.

\subsection{Simulation results}

\begin{figure*}[t!]
\noindent \includegraphics[width=16.9cm]{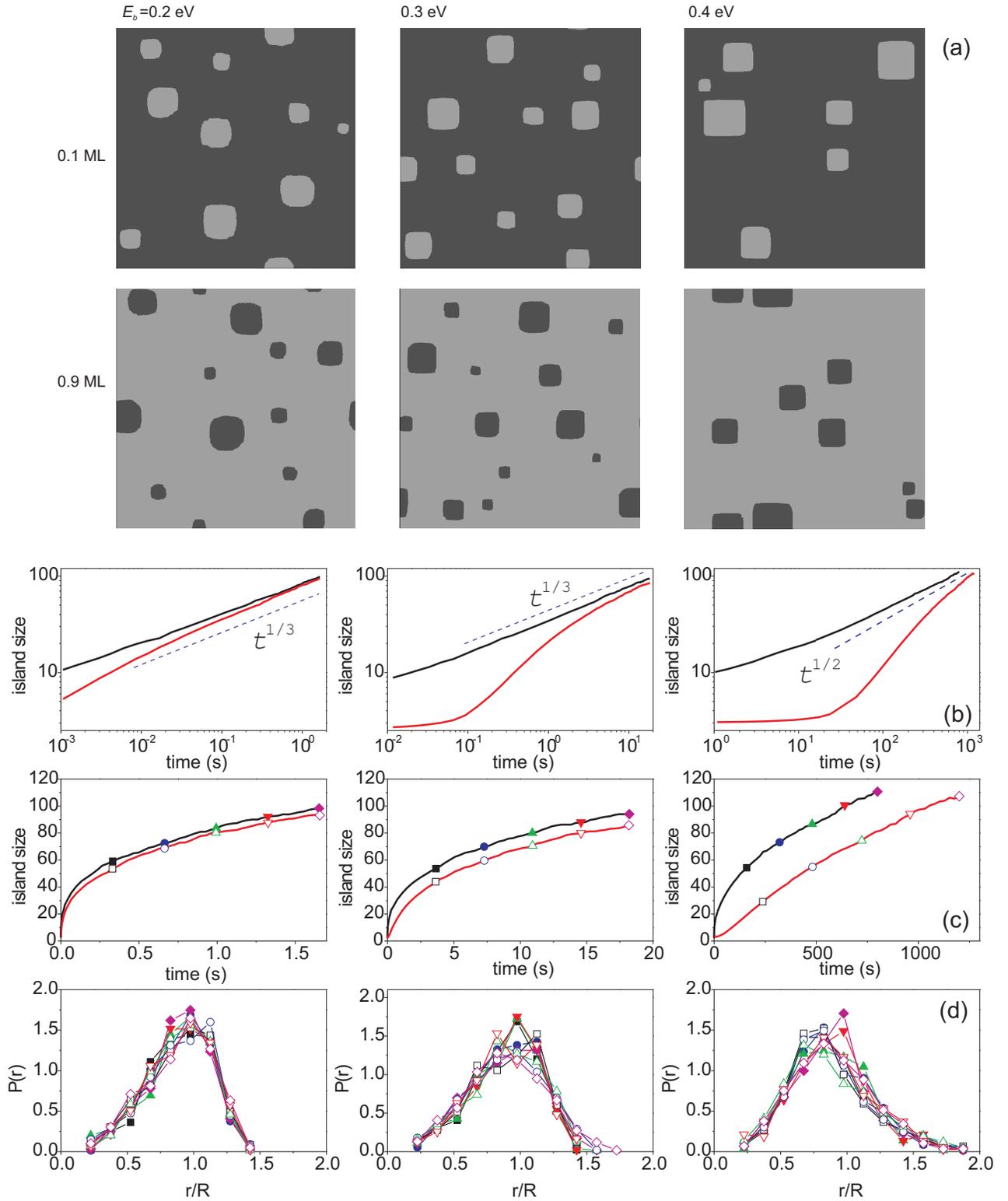}
\caption{Results of kinetic Monte Carlo simulations: (a) snapshots
of the 1000$\times$1000 simulation cells at the end of the
simulations, (b) and (c) time dependence of the average island size
in logarithmic and linear scales, and (d) the island size
distributions. The gray levels in the snapshots vary from black to
white as the surface height increases. Different columns show
results for different bond energies $E_{b}$, with the temperature
fixed at $T=400$~K. The size distributions are obtained at the time
moments marked in (c) by the corresponding symbols.} \label{kMC}
\end{figure*}

Snapshots of the simulated system at the end of a simulation are
presented in Fig.\ \ref{kMC}(a). As the bond energy $E_{b}$ is
increased (from left to right), the island shape continuously
transforms from more circular to almost square. Since faceting
transitions are absent in 2D systems, we refer to the almost square
islands as faceted in order to stress the qualitative shape
difference at small and large bond energies. Apart from the change
in shape, the equilibrium density of adatoms between islands
exponentially decreases as the bond energy increases.

Figures \ref{kMC}(b) and (c) show time variations of average island
diameters $2R(t)$ in logarithmic and linear scales, respectively.
The determination of an average island size is described in Sec.\
\ref {subsec:analysis}. At small bond energies (left column in Fig.\
\ref{kMC}), the process of Ostwald ripening follows the
Lifshitz--Slyozov law $R(t)\propto t^{1/3}$. As the bond energy
increases, the coarsening law for advacancy islands deviates from
that for adatom islands and from the expected $t^{1/3}$ law. At
large bond energies (right column in Fig.\ \ref {kMC}), the
coarsening behavior of advacancy islands is qualitatively different
and close to a linear dependence, in a wide range of island sizes.
The coarsening of adatom islands also notably deviates from the
Lifshitz--Slyozov law. The attachment-limited asymptotic $t^{1/2}$
can be inferred from the figure, but it is not really reached.

Figure \ref{kMC}(d) shows the island size distributions at different
times. The uniformly spaced time instances are marked on the curves
in Fig.\ \ref{kMC}(c) by the same symbols as used for the
corresponding size distributions. The distributions are scaled by
the average size $R(t)$: instead of the probability $P(r)$, we plot
$RP(r)$ versus $r/R$. The scaled distributions do not change in time
even at large bond energies, where the average island sizes do not
show a power law behavior. The island size distribution notably
changes with increasing bond energy, Fig.\ \ref{kMC}(d). The
distribution develops a tail extended to $2R$, while at smaller bond
energies it is limited to $1.5R$.

\subsection{Analysis of the simulation data}

\label{subsec:analysis}We obtain the sizes of all islands in the
simulated system by using an algorithm\cite{hoshen76} that allows to
count all topologically connected clusters in the system. At large
bond energies, we average the radii $r_{n}=\sqrt{n/\pi }$ (where $n$
is the number of atoms in a cluster) of all islands, excluding
individual adatoms from the distribution. In the case of small bond
energies we find that, besides monomers, a notable amounts of
transient dimers, trimers, etc.\ are present in the simulated
system. Their densities quickly decrease with increasing number of
atoms in the cluster and they are well separated from the
distribution of the large clusters. If these small clusters are
included in the island size distribution when calculating average
radius $R$, we obtain unreasonable time dependencies $R(t)$. Hence,
we calculate the averages taking into account islands of at least 6
atoms.

\begin{figure}[tbh]
\noindent \includegraphics[width=\columnwidth]{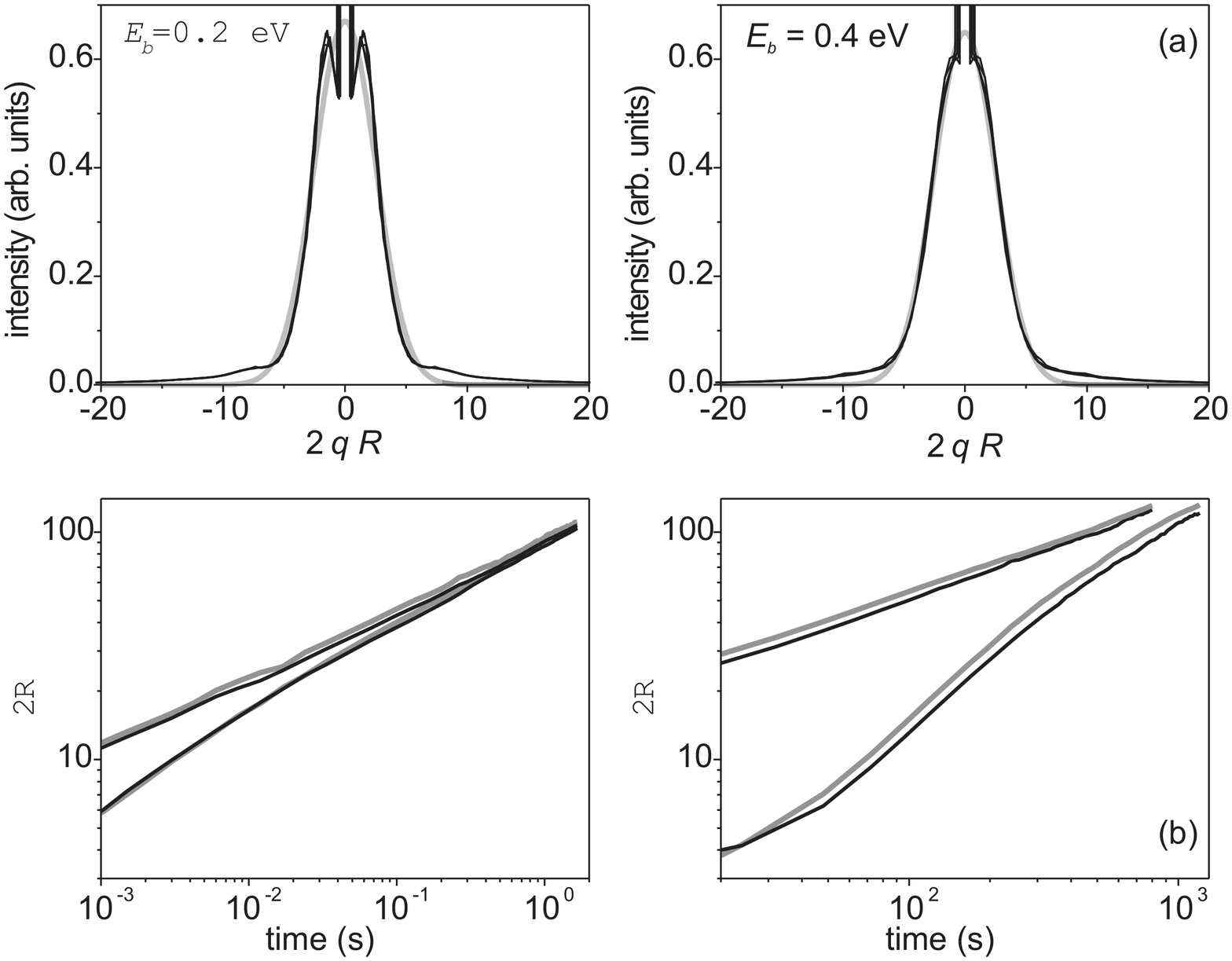}
\caption{(a) Diffraction peaks obtained from the Monte Carlo
simulation results (the gray curves are Gaussian fits). (b) Time
dependencies of the average island sizes obtained from the numbers
of atoms in the clusters (black curves) and from the widths of the
diffraction peaks (gray curves).} \label{MCanalysis}
\end{figure}

We also use the Monte Carlo simulations to verify the average island
size determination in diffraction studies. In a diffraction
experiment, one has access to the peak profile only and obtains the
average size from its width. Using the island distribution obtained
in the simulation and calculating the peak profiles, we can compare
the average sizes obtained from the real space and the reciprocal
space distributions. The diffraction peaks (structure factors)
obtained from the simulations are present in Fig.\
\ref{MCanalysis}(a). We consider the anti-Bragg condition
(subsequent atomic layers contribute to the scattering function with
a phase shift of $\pi $) and obtain two-dimensional intensity
distributions $I(q_{x},q_{y})$ from Fourier transformation of $\exp
[i\pi h(x,y)]$. Here an integer function $h(x,y)$ is the surface
height. Then, we take into account that in a diffraction experiment,
the scattered intensity is usually collected by a wide open detector
that integrates over one of the components of the scattering vector
$\mathbf{q}$.\cite{braun04prb} Hence, we integrate the distributions
$I(q_{x},q_{y})$ over one of the components of the scattering
vector, either $q_{x}$ or $q_{y}$. The resulting diffraction peaks
$I(q)$ are presented in Fig.\ \ref{MCanalysis}(a). The peaks
corresponding to different time moments [the same time moments as in
Fig.\ \ref{kMC}(d)] coincide after the wave vectors $q$ are scaled
by the average island size. Kinetic scaling is thus confirmed. The
shapes of the peaks depend on the bond energy $E_{b}$, thus showing
that the island size distribution and the correlations between
islands change.

The quantity most commonly measured in a diffraction experiment is
the full width at half maximum (FWHM)\ of a peak obtained by an
appropriate fit. Considering islands of linear size $2R$, one
obtains a structure factor $\sin ^{2}(qR)/\sin ^{2}(qa)$, which can
be approximated by $\exp (-q^{2}R^{2}/\pi )$.\cite{warren:book69}
Here, $a$ is the lattice spacing. We obtain the average size $2R$ by
fitting the peaks to this Gaussian function, despite the peaks are
not Gaussian, especially for small bond energies. Figure
\ref{MCanalysis}(b) compares these sizes with the ones obtained from
the real-space island size analysis described above. The values are
in good agreement, thus confirming that the average quantities can
be obtained from the diffraction peak widths even if the profiles
deviate notably from Gaussian.

\section{Coarsening equations}

\subsection{The Becker--D\"{o}ring equations for the 3D problem}

The process of Ostwald ripening can be described by two alternative
approaches, either in terms of a continuous function $f(r)$\
representing the number density of clusters of radius $r$, or in
terms of discrete numbers $c_{n}$ representing the densities of
clusters containing $n$ atoms ($n$mers). The first approach was
employed by Lifshitz and Slyozov\cite {lifshitzslezov61} and
Wagner.\cite{wagner61} The equations for discrete quantities $c_{n}$
were first formulated by Becker and D\"{o}ring\cite
{BeckerDoering35} and ever since form the basis of nucleation
theory.\cite {frenkel46,LewisAnderson78} Closely related equations,
the rate equations, were used in the description of crystal
growth.\cite {zinsmeister66,stoyanov81,venables84} They contain an
additional deposition term, while the detachment process is not
essential and the corresponding terms in the equations are
frequently omitted. Similar discrete equations for the Ostwald
ripening process were introduced under the names of microscopic
continuity equations,\cite{dadyburjor77,bhakta95} population balance
equations, \cite{madras02,madras03,madras05} or rate equation
approach.\cite{xia98} Mathematical aspects of the relationship
between the discrete and the continuous equations were also
considered.\cite {penrose97,collet02} The aim of the present section
is to link the discrete and continuous approaches and obtain
equations that can be used for a numerical study of the Ostwald
ripening process.

The number of atoms $n$ in a cluster increases or decreases by one
when an atom is attached to the cluster or detached from it. Let
$J_{n}$ be the net rate of transformation of $n$mers into
$(n+1)$mers. The number $c_{n}$ of $n$mers increases due to the
transformation of $(n-1)$mers into $n$mers and decreases because of
the transformation of $n$mers into $(n+1)$mers:
\begin{equation}
dc_{n}/dt=J_{n-1}-J_{n}.  \label{eq1}
\end{equation}
This equation is valid for $n\geq 2$. The equation describing the number of
monomers $c_{1}$ is obtained by requiring that the total number of atoms in
the system
\begin{equation}
N=\sum_{n=1}^{\infty }nc_{n}  \label{eq2}
\end{equation}
does not change in time. The condition $dN/dt=0$ gives, after substitution
of Eqs.\ (\ref{eq1}) and rearrangement of the terms,
\begin{equation}
dc_{1}/dt=-2J_{1}-\sum_{n=2}^{\infty }J_{n}.  \label{eq3}
\end{equation}
This equation takes into account that each transformation of an
$n$mer into an $(n+1)$mer decreases the number of monomers by one,
except in the case $n=1$, where two monomers form a dimer.

The net rate $J_{n}$ is a result of two processes. First, an $n$mer
catches a monomer. The rate of this process is proportional to the
densities of the $n$mers and the monomers and can be written as
$a_{n}c_{1}c_{n}$, where $a_{n} $ is a time-independent coefficient
that remains to be determined. The second process is a spontaneous
detachment of a monomer from a $(n+1)$mer. It is proportional to the
density of $(n+1)$mers solely and can be written as $b_{n}c_{n+1}$,
where $b_{n}$ is another time-independent coefficient to be
specified. Hence, we obtain
\begin{equation}
J_{n}=a_{n}c_{1}c_{n}-b_{n}c_{n+1}.  \label{eq4}
\end{equation}
Equations (\ref{eq1}),\ (\ref{eq3}), and (\ref{eq4}) are the
Becker--D\"{o}ring equations.

If the time limiting process is the adatom diffusion between
clusters, the attachment and detachment coefficients $a_{n}$ and
$b_{n}$ for the 3D\ problem are calculated, for large $n,$ as
follows. The cluster of $n$ atoms is considered as a sphere of
radius $r_{n}$, so that $n=4\pi r_{n}^{3}/3$. To calculate the
attachment coefficient, we solve the steady-state diffusion equation
$\nabla ^{2}c(r)=0$ with two boundary conditions:\ the concentration
of the monomers far away from the cluster is equal to their mean
concentration, $c(r)\left| _{r=\infty }\right. =c_{1}$, while the
concentration of the monomers at the cluster surface is zero,
$c(r)\left| _{r=r_{n}}\right. =0$, since the monomers are attached
to the cluster as soon as they reach it. The solution is
$c(r)=(1-r_{n}/r)c_{1}$. The total atom flux at the cluster surface
\begin{equation}
j_{n}=4\pi r_{n}^{2}D\nabla c(r)\left| _{r=r_{n}}\right. ,  \label{eq4b}
\end{equation}
where $D$ is the diffusion coefficient of the monomers, is equal to $4\pi
Dr_{n}c_{1}$, and hence the attachment coefficient is
\begin{equation}
a_{n}=4\pi Dr_{n}.  \label{eq4a}
\end{equation}

The detachment coefficient is calculated assuming that the
concentration of the monomers at the cluster surface is equal to the
equilibrium monomer concentration $c_{n\mathrm{eq}}$, while there is
an ideal sink for monomers at infinity, $c(r)\left| _{r=\infty
}\right. =0$. The solution of the steady-state diffusion equation
with these boundary conditions is $c(r)=c_{n\mathrm{eq}}r_{n}/r$,
and the corresponding detachment flux of the monomers is
$b_{n+1}=4\pi Dr_{n}c_{n\mathrm{eq}}$. Here we take into account
that this flux refers to the detachment from the $(n+1)$mer. The
ratio of the detachment and the attachment coefficients is then
\begin{equation}
b_{n+1}/a_{n}=c_{n\mathrm{eq}}.  \label{eq5}
\end{equation}
The equilibrium density of monomers at the surface of a cluster is given by
the Gibbs--Thomson formula
\begin{equation}
c_{n\mathrm{eq}}=c_{\infty \mathrm{eq}}\exp (\gamma /r_{n})\approx c_{\infty
\mathrm{eq}}(1+\gamma /r_{n}),  \label{eq6}
\end{equation}
where $\gamma $ is a constant proportional to the surface tension.
The explicit expression for $\gamma $ is given in the next section.
A correction to Eq.\ (\ref{eq6}) for small clusters consisting of
very few atoms, that is important in the theory of nucleation, is
not essential for the Ostwald ripening problem. Then, equations
(\ref{eq1})--(\ref{eq6}) give a complete set of equations that
describe the process of Ostwald ripening.

When clusters are large enough, $n$ can be treated as a continuous
variable. Let us verify that the continuous equations derived from
the set of equations above are the Lifshitz--Slyozov equations. The
cluster size distribution function $f(r,t)$ is defined so that
$f(r,t)dr$ is the number of clusters per unit volume in an interval
from $r$ to $r+dr$. Then, $f(r,t)dr=c_{n}(t)dn$ and, keeping in mind
that $n=4\pi r^{3}/3$, we obtain $f(r,t)=4\pi r^{2}c_{n}(t)$. The
mass conservation law (\ref{eq2}) can be rewritten, by separating
monomers and larger clusters, as
\begin{equation}
c_{1}(t)+\frac{4\pi }{3}\int_{0}^{\infty }r^{3}f(r,t)dr=N=\mathrm{const}.
\label{eq7}
\end{equation}
The finite-difference equation (\ref{eq1}) transforms into the continuity
equation
\begin{equation}
\partial f/dt+\partial J/\partial r=0.  \label{eq8}
\end{equation}
To calculate the flux in the cluster size space $J(r,t)$, one can neglect
the difference between $c_{n}$ and $c_{n+1}$ in Eq.\ (\ref{eq4}). Then,
substituting Eqs.\ (\ref{eq5}) and (\ref{eq6}), one obtains
\begin{equation}
J(r,t)=\frac{D}{r}(c_{1}-c_{\infty \mathrm{eq}}-\frac{\gamma c_{\infty
\mathrm{eq}}}{r})f.  \label{eq9}
\end{equation}
Equations (\ref{eq7})--(\ref{eq9}) coincide with the
Lifshitz--Slyozov equations.\cite{lifshitzslezov61}

\begin{figure}[tbh]
\noindent \includegraphics[width=\columnwidth]{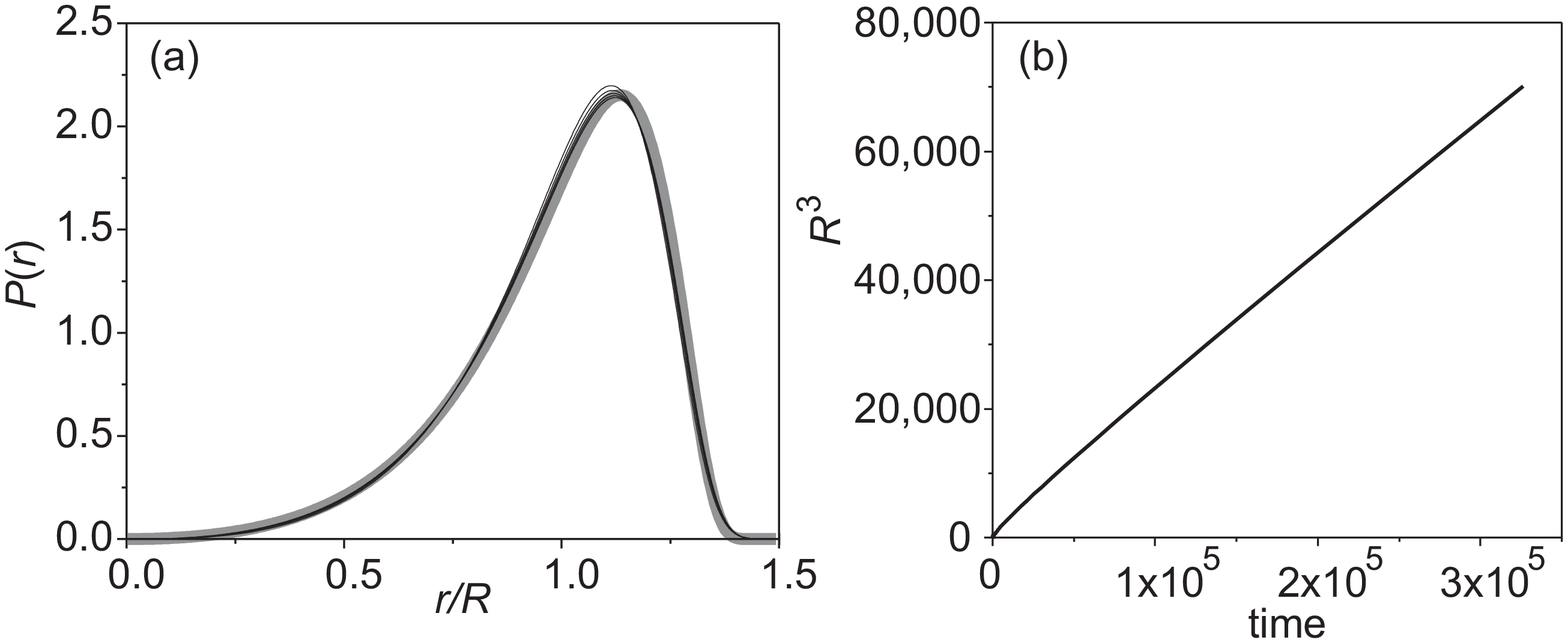}
\caption{ (a) The cluster size distribution obtained by numerical solution
of the Becker--D\"oring equations at different times (thin black lines, the
lines closer to the gray line correspond to later times) and the analytical
solution by Lifshitz and Slyozov (thick gray line). (b) The time dependency
of $R^{3}$. A linear asymptotic is evident from the plot.}
\label{analytical}
\end{figure}

As an example, we compare in Fig.\ \ref{analytical} numerical
solutions of the ordinary differential equations
(\ref{eq1})--(\ref{eq6}) with the analytical
result.\cite{lifshitzslezov61} To solve the Becker--D\"oring system,
we employ a second-order Rosenbrock method, which is essentially
based on a Pade-approximation of the transition operator (see, e.g.,
Ref.\ \onlinecite{hairer96}). A version of this
method\cite{levykin1996} that fits well to stiff systems of
differential-algebraic equations was used. Practically, we solve a
set of up to one million ordinary differential equations on a
personal computer. The solutions in Fig.\ \ref{analytical} are
obtained by taking $\gamma =5$ and, as the initial condition at
$t=0,$ only monomers with the initial supersaturation
$c_{1}/c_{\infty \mathrm{eq}}=10^{5}$. The figure shows that the
numerical solutions asymptotically converge to the analytical
formula, which validates our approach.

\subsection{Attachment and detachment coefficients}

\label{subsec:AttachDetach}

Equation (\ref{eq5}) can be derived in a more general form that will be
useful for the considerations below. In equilibrium, all fluxes $J_{n}$ are
identically equal to zero. Then, denoting by $C_{n}$ the equilibrium
concentrations of the $n$mers, we have from Eq.\ (\ref{eq4})
\begin{equation}
b_{n+1}/a_{n}=C_{1}C_{n}/C_{n+1}.  \label{eq10}
\end{equation}
The equilibrium concentrations calculated in the framework of equilibrium
thermodynamics are\cite{walton62}
\begin{equation}
C_{n}=C_{1}^{n}\exp [-(E_{n}-nE_{1})/k_{\mathrm{B}}T],  \label{eq11}
\end{equation}
where $E_{n}$ is the energy of an $n$mer and $E_{1}$ is the energy of a
monomer. This relation can be treated as the mass action law for the
equilibrium between $n$mers and monomers, $C_{n}\leftrightarrows nC_{1}$.
Substitution into Eq.\ (\ref{eq10}) gives
\begin{equation}
b_{n+1}/a_{n}=c_{\infty \mathrm{eq}}\exp [(E_{n+1}-E_{n})/k_{\mathrm{B}}T],
\label{eq12}
\end{equation}
where $c_{\infty \mathrm{eq}}=\exp (-E_{1}/kT)$ is the concentration
of monomers that are in equilibrium with an infinite cluster. For
spherical clusters, Eq.\ (\ref{eq12}) reduces to the Gibbs--Thomson
formula. The energy of a spherical cluster is $E_{n}=4\pi
r^{2}\sigma $, where $\sigma $ is the surface tension, with the
radius $r$ defined by $nv=4\pi r^{3}/3$, where $v=a^{3}$ is the
volume per atom. The radius increase due to the attachment of an
atom to a $n$mer is given by $v=4\pi r^{2}\Delta r$. The change of
the energy due to the attachment of a single atom is
$E_{n+1}-E_{n}=8\pi \sigma r\Delta r=2v\sigma /r$. Thus, we arrive
at Eq.\ (\ref{eq6}) with $\gamma =2v\sigma /k_{\mathrm{B}}T$. A
similar calculation for the 2D case gives $\gamma =s\sigma
/k_{\mathrm{B}}T$, where $s$ is the area per atom.

\begin{figure}[tbh]
\noindent \includegraphics[width=0.8\columnwidth]{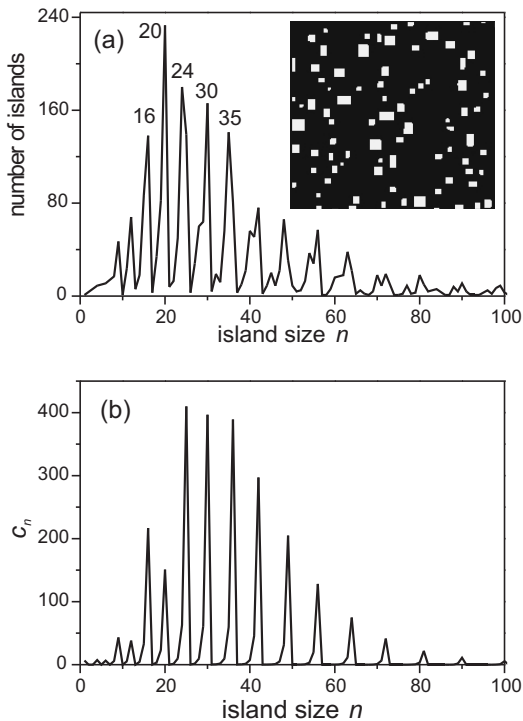}
\caption{The island size distribution of faceted islands obtained in
the kinetic Monte Carlo simulations (a) and by numerical solution of
the Becker--D\"oring equations (b). The strong preference of magic
island sizes is obvious.} \label{faceted}
\end{figure}

Equation (\ref{eq12}) is more general than the Gibbs--Thomson
formula and can be used in situations when the latter is not
applicable. Figure \ref {faceted}(a)\ presents the island size
distribution obtained in our kinetic Monte Carlo simulations at an
early stage of coarsening for the largest bond energy we have
studied, $E_{b}=0.4$ eV. The distribution is not smooth but consists
of peaks at `magic' island sizes corresponding to a product of two
close integers, like $30=6\times 5$. Accordingly, the insert in the
figure shows that the islands are mainly rectangles with an aspect
ratio close to 1. The origin of such a distribution is evident: when
an island consisting, for example, of 30 atoms, grows by one atom,
its energy increases by $2E_{b}$, while further growth to 36 atoms
does not change its energy at all. Thus, we solve the
Becker--D\"{o}ring equations with the energy of a 2D island of $n$
atoms calculated as follows. First, we find the largest square that
still contains fewer atoms than $n$. Then, we add, as long as the
number of atoms does not exceed $n$, rows of atoms to the side of
the square. The last row may be incomplete. The number of broken
bonds for such an island is calculated. Figure \ref{faceted}(b)\
presents a numerical solution of the Becker--D\"{o}ring equations
with the island energies $E_{n}$ thus calculated and the
attachment--detachment coefficient ratio given by Eq.\ (\ref{eq12}).
The approximation for $a_{n}$ appropriate for the 2D case is given
below in Sec.\ \ref{coars2D}. The size distribution closely
reproduces the one obtained in the Monte Carlo simulations: squared
or rectangular (with aspect ratio close to 1) islands are discrete
barriers to be overcome, while the filling of an atomic row does not
change the island energy and proceeds relatively fast. This example
shows that Eq.\ (\ref{eq12}) can be used when the island energy
$E_{n}$ is known but is not described simply by the surface tension,
and the Gibbs--Thomson formula is not applicable.

\subsection{Coarsening equations in two dimensions}

\label{coars2D}

The Becker--D\"{o}ring equations (\ref{eq1})--(\ref{eq4}) and the
equation (\ref{eq12}) for the ratio of the coefficients
$b_{n+1}/a_{n}$ do not depend on the dimensionality of the system
and can be applied to both 2D and 3D problems. (It may be worth to
note that the radius $r_{n}$ entering the Gibbs--Thomson law is
expressed differently through $n$ in the 2D and 3D cases.) The only
formula that has to be reconsidered is expression (\ref {eq4a}) for
the attachment coefficients $a_{n}$, since it is based on the
solution of the 3D diffusion equation. The solution of the 2D
diffusion equation behaves as $c(r)\propto \ln r$ and the boundary
condition $c(r)\left| _{r=\infty }\right. =c_{1}$ cannot be imposed.
A simple approximation is to place this condition at a finite
distance $l$, given by an average distance between the islands.\cite
{chakraverty67,rogers89,ardell90,bhakta95,hausser2005} Then, in the
case of diffusion-limited kinetics, the attachment coefficient
$a_{n}$ does not depend on $n$ and is proportional to $(\ln
l)^{-1}$. Proceeding to the continuous distribution function, one
arrives at Eq.\ (\ref{eq9}), with the conservation law (\ref{eq7})
rewritten for the 2D case. The coarsening equations are solved
analytically in this case.\cite {hillert65,rogers89,ardell90}

A self-consistent description of two-dimensional diffusion can be
obtained by taking into account its screening by the island
distribution.\cite{marqusee84} A solution of the 2D screened
diffusion equation, satisfying the boundary conditions $c(r)\left|
_{r=\infty }\right. =c_{1}$ and $c(r)\left| _{r=r_{n}}\right. =0$,
is $c(r)=c_{1}[1-K_{0}(r/\xi )/K_{0}(r_{n}/\xi )]$, where $K_{0}(x)$
is the zeroth modified Bessel function and $\xi $ is the screening
length that remains to be defined. Then, one obtains the attachment
coefficient
\begin{equation}
a_{n}=D\mathcal{K}(r_{n}/\xi ),  \label{eq14}
\end{equation}
where
\begin{equation}
\mathcal{K}(x)=2\pi xK_{1}(x)/K_{0}(x)  \label{eq15}
\end{equation}
and $K_{1}(x)$ is the first modified Bessel function. The self-consistency
condition for the screening length $\xi $ is\cite{marqusee84}
\begin{equation}
\xi ^{-1}=\int_{0}^{\infty }\mathcal{K}(r/\xi )f(r,t)dr.  \label{eq16}
\end{equation}

Expressions very similar to Eqs.\ (\ref{eq14}) and (\ref{eq15}) are used in
studies of crystal growth from the gas phase\cite
{LewisAnderson78,stoyanov81,venables84}, with one essential difference:\ for
the latter problem, the length $\xi $ is the mean diffusion length of an
adatom on the surface before its reevaporation. It is a well-defined
time-independent constant, so that no self-consistency condition is involved.

In the case of attachment-limited kinetics, the boundary condition
for the concentration field $c(r)$ at the island surface is the
absence of the flux, $\nabla c\left| _{r=r_{n}}\right. =0$, which
gives a constant solution, $c(r)=c_{1}$. Then, the attachment
coefficient is
\begin{equation}
a_{n}=2\pi Kr_{n},  \label{eq16b}
\end{equation}
where $K$ is the attachment coefficient. The result is independent
of screening effect in this case. The same expression is obtained in
the approximation of a constant screening distance equal to the mean
distance between islands. \cite
{chakraverty67,rogers89,ardell90,bhakta95,hausser2005}

\subsection{Coarsening equations for advacancy islands}

In our Monte Carlo simulations, a step edge barrier is absent and an
atom detaching from a vacancy island ascends to the higher terrace.
The vacancy island size increases by a vacancy at the same time. The
coarsening proceeds by exchange of adatoms between vacancy islands
and can be described by equations similar to the Becker--D\"{o}ring
equations. Let us denote by $g(t) $ the concentration of adatoms,
while $c_{n}$ are the concentrations of 2D islands of $n$ vacancies.
Then, the continuity equation (\ref{eq1}) for the density of
clusters $c_{n}(t)$ remains unchanged. The fluxes $J_{n}$ in these
equations describe two processes. The first process is the
spontaneous emission of an adatom. Its rate is proportional to the
density of $n$mers. The second process is an absorption of an adatom
by the vacancy type\ $(n+1)$mer, which gives rise to a $n$mer. Its
rate is proportional to the density $g$ of adatoms and the density
of $(n+1)$mers, so that
\begin{equation}
J_{n}=b_{n}c_{n}-a_{n+1}gc_{n+1}.  \label{eq17}
\end{equation}
The annihilation of an atom and a single vacancy is described by the
flux $J_{0}=-a_{1}gc_{1}$. Then, the set of equations (\ref{eq1}) is
valid for $n\geq 1$. The creation of an adatom--vacancy pair from a
flat surface is prohibited in our model.

Since the growth of a vacancy cluster by one vacancy is accompanied with the
emission of one adatom, the conserved total amount of atoms in the system is
given by
\begin{equation}
N=\sum_{n=1}^{\infty }nJ_{n}-g,  \label{eq18}
\end{equation}
which replaces Eq.\ (\ref{eq2}). By differentiating this equation
with respect to time and rearranging the terms, we obtain from
$dN/dt=0$ an equation for the time variation of the adatom density:
\begin{equation}
dg/dt=\sum_{n=0}^{\infty }J_{n}.  \label{eq19}
\end{equation}

The mass action law now has to be written for an equilibrium between
an advacancy island and adatoms that annihilate,
$C_{n}+ng\leftrightarrows 0$. Hence, instead of Eq.\ (\ref{eq11}) we
have
\begin{equation}
C_{n}g^{n}=\exp [-(E_{n}+nE_{1})/k_{\mathrm{B}}T].  \label{eq19a}
\end{equation}
The requirement of zero fluxes at equilibrium gives rise to the detailed
balance condition
\begin{equation}
b_{n}/a_{n+1}=c_{\infty \mathrm{eq}}\exp [-(E_{n+1}-E_{n})/k_{\mathrm{B}}T]
\label{eq20}
\end{equation}
that differs from Eq.\ (\ref{eq12}) by the sign in the exponent. For
circular islands, the same calculation as above leads to the Gibbs--Thomson
formula (\ref{eq6}) with negative $\gamma $, which corresponds to a negative
curvature of the vacancy island surface.

\subsection{Solutions of the coarsening equations}

\begin{figure}[tbh]
\noindent \includegraphics[width=\columnwidth]{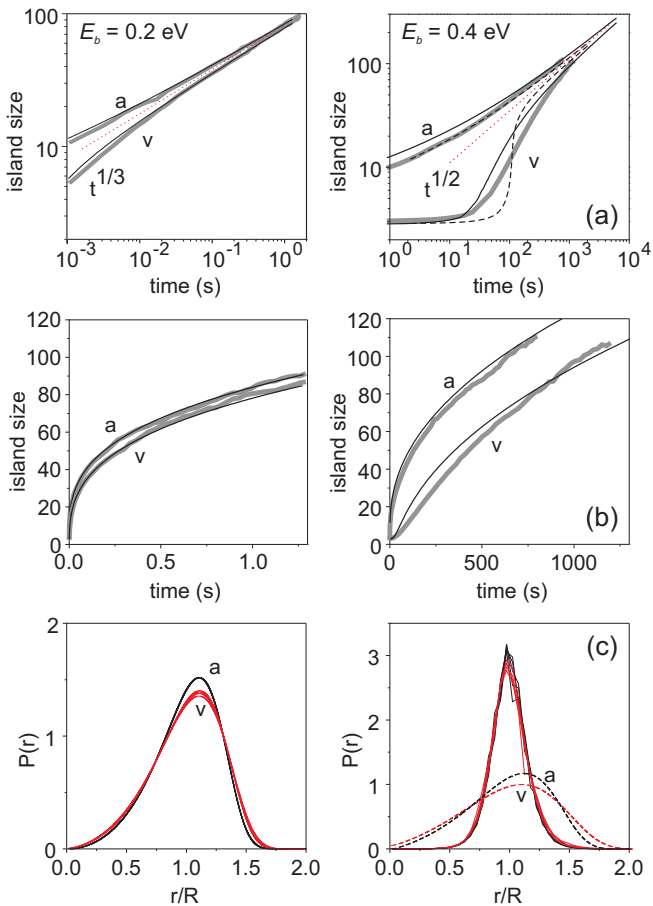}
\caption{The results of numerical solution of the Becker--D\"oring
equations: time dependencies of the average island sizes in
logarithmic (a) and linear (b) scales and the island size
distributions (c). The left column presents calculations for the
bond energy $E_b=0.2$ eV with diffusion-limited kinetics, while the
right column shows the results for the bond energy $E_b=0.42$ eV
with attachment-limited kinetics. The solutions of the
Becker--D\"oring equations are shown in (a) and (b) by black lines,
and the results of the kinetic Monte Carlo simulations by gray
lines. Symbols ``a'' and ``v'' on the plots denote the results for
adatom and advacancy islands, respectively. Full lines in the right
column show the calculations for the discrete island energies with
`magic' sizes, while the broken lines are calculations for the
continuous Gibbs-Thomson chemical potential.} \label{BDresults}
\end{figure}

Figure \ref{BDresults} presents the results of the numerical
solution of the Becker--D\"{o}ring equations for adatom and
advacancy islands. With the aim to quantitatively compare the
solutions with the results of kinetic Monte Carlo simulations in the
whole time interval, we obtain the average island sizes in the same
way as in the simulations, and use the same initial conditions.
Namely, the average island sizes are calculated taking into account
the islands containing at least 6 atoms, for the reasons discussed
in Sec.\ \ref{subsec:analysis}. The initial random adatom
distribution with the coverage 0.1 ML contains not only monomers,
but also dimers, trimers, etc., the densities of which quickly
decrease with the increasing number of atoms in the cluster. By
simple statistical analysis of the initial distribution in kinetic
Monte Carlo simulations, we find that at $t=0$, $c_{n}\approx
c_{1}\times 10^{(n-1)/2}$. This distribution was used as the initial
condition for the Becker--D\"{o}ring equations. The initial
conditions are essential only at the initial stages of coarsening.
The results of the calculations do not depend on the initial monomer
concentration $c_{1}$, as long as the initial supersaturation
$c_{1}(t=0)/c_{\infty \mathrm{eq}}$ is much larger than unity. The
time scales of the solutions are adjusted to these of the Monte
Carlo simulations.

The case of small bond energies (left column in Fig.\
\ref{BDresults}) is well described by the 2D diffusion limited
kinetics with screening (\ref {eq14}) and the ratio of the
detachment and the attachment coefficients given by the
Gibbs--Thomson formula (\ref{eq6}). Calculations in the left column
of Fig.\ \ref{BDresults} are made with $\gamma =3.7$. The solutions
of the Becker--D\"{o}ring equations (black lines) are in a good
agreement with the results of the kinetic Monte Carlo simulations
(gray lines), that are repeated from Fig.\ \ref{kMC}. The coarsening
laws for adatom and advacancy islands almost coincide and quickly
reach the Lifshitz--Slyozov $t^{1/3}$ asymptotic. The island size
distributions, Fig.\ \ref{BDresults}(c), also almost coincide for
adatom and advacancy islands, possess kinetic scaling, and agree
well with these obtained in the kinetic Monte Carlo simulations,
cf.\ Fig.\ \ref{kMC}(d).

For large bond energies (right column in Fig.\ \ref{BDresults}), the
calculations are performed with attachment-limited kinetics, Eq.\
(\ref {eq16b}), since the kinetic Monte Carlo simulations point to
the Wagner's $t^{1/2}$ asymptotic. We compare the discrete
distribution of the island energies that takes into account the
`magic' island sizes as described in Sec.\ \ref
{subsec:AttachDetach} (full black lines) with the continuous one,
given by the Gibbs--Thomson formula (broken lines). The relationship
between the discrete and the continuous models is established by
calculating the energy of a square island and a circular one with
the same number of atoms:\ $E_{b}/k_{\mathrm{B}}T=\sqrt{\pi }\gamma
/2$. The calculations are performed for $\gamma =7$. The effect of
magic sizes is slightly different for adatom and advacancy islands.
For adatom islands, the detachment coefficients $b_{n} $ given by
Eq.\ (\ref{eq12}) are exceptionally large for $n=m+1$, where $m$ is
a magic number. Thus, a monomer that has attached to a magic island
detaches again with a high probability. For advacancy islands, the
detachment coefficients $b_{m}$ for magic islands are exceptionally
small, so that the detachment of an atom from a vacancy island (this
atom becomes an adatom on the higher level) proceeds at a small
rate. Both processes make each magic size a trap for further island
growth, giving rise to the discrete island size distribution peaked
at the magic sizes, shown in Fig.\ \ref{faceted}. The island size
distributions presented in Fig.\ \ref{BDresults}(c) for this case
are obtained by averaging over finite ranges of the sizes, in the
same way as done in the kinetic Monte Carlo simulations.

The time dependence of the average island sizes obtained for
coarsening through the sequence of magic islands (full black lines
in right column of Fig.\ \ref {BDresults}) are in good agreement
with the results of kinetic Monte Carlo simulations (gray lines).
For vacancy islands, the continuous island size distribution with
the Gibbs--Thomson formula gives rise to a notably different
coarsening behavior (broken lines), with a very fast increase of the
island sizes in an intermediate range. The island size distributions
obtained in the discrete (with magic sizes) and the continuous
models are also notably different, see Fig.\ \ref{BDresults}(c). The
distribution obtained in the discrete model is symmetric with
respect to the maximum, similar to the one obtained in the Monte
Carlo simulations, but notably narrower, cf.\ Fig.\ \ref{kMC}(d). It
is worth to note that the distribution scaled by the average island
size does not change in time and is the same for the adatom and
advacancy islands, despite the time evolutions of the average island
sizes not coinciding and not following a power law. In other words,
the solution of the Becker--D\"{o}ring equation obeys kinetic
scaling in the sense that the island size distribution is a function
of $r/R(t)$ that does not depend on time. However, $R(t)$ is not
described by a power law. The continuous model gives a much broader
and asymmetric island size distribution, shown by broken lines in
Fig.\ \ref {BDresults}(c). The broken-bond counting scheme described
in Sec.\ \ref{subsec:AttachDetach} adequately represents the
energies $E_{n}$ of small islands and quantitatively describes the
island size distribution at the initial stage of coarsening, see
Fig.\ \ref{faceted}. However, for larger islands it\ oversimplifies
the island energy distribution and gives rise to a more narrow
distribution than found in the simulations. A better model for the
island energies $E_{n}$ is needed to describe this distribution
correctly.

To summarize this section, we show that the Ostwald ripening
kinetics can be described as an initial value problem for ordinary
differential equations (\ref{eq1})--(\ref{eq6}) that can be solved
by standard numerical methods. This approach can be applied to
various coarsening problems by replacing the Gibbs--Thomson formula
(\ref{eq6}) with Eqs.\ (\ref{eq12}), (\ref{eq20}) that admit any
dependence of the island energy $E_{n}$ on the number of atoms $n$
in it. The alternative approach, a numerical implementation of the
integro-differential equations (\ref{eq7})--(\ref{eq9}),\cite
{venzl83,carrillo04} seems much more difficult.

\section{Conclusions}

Our kinetic Monte Carlo simulations show that the Ostwald ripening
of 2D islands qualitatively changes with increasing bond energy (or
decreasing temperature). The islands become faceted and the
coarsening proceeds through a sequence of magic sizes. The
Gibbs-Thomson chemical potential is not applicable and the
detachment of monomers from islands is governed by the discrete
energies of the islands. The coarsening is diffusion limited at
small bond energies and becomes attachment limited at large bond
energies. In this latter case, Wagner's asymptotic law is reached
only after a very long transient regime.

We show that the Becker--D\"{o}ring equations of nucleation kinetics
are well suited to study the process of Ostwald ripening. Two- and
three-dimensional coarsening processes with diverse limiting
mechanisms can be simulated by solving a system of ordinary
differential equations. Concentrations of clusters of all sizes,
from monomers to ones consisting of millions of atoms, can be traced
simultaneously. The calculations are not necessarily based on the
Gibbs--Thomson formula but adopt any continuous or singular
dependence of the cluster energy on the number of atoms in it. This
approach can be applied to a wide range of coarsening problems for
two- and three-dimensional islands on a surface.

\acknowledgments

We thank D.\ Wolf and H.\ M\"uller-Krumbhaar for stimulating
discussions, and A.\ I.\ Levykin for his advice. Financial support
from RFBR (Grant N 06-01-00498) is acknowledged.


\end{document}